\documentclass{aastex61}
\received{\today}
\revised{}
\accepted{}
\submitjournal{ApJ}
\shorttitle{Poloidal fields in accretion disks}
\shortauthors{Samadi \& Abbassi}

\begin{document}

\title{Anchoring polar magnetic field in a stationary thick accretion disk}
%\correspondingauthor
\author{Maryam Samadi}
\email{samadimojarad@um.ac.ir}

%\author[0000-0002-0786-7307]{Maryam Samadi}
\affil{Department of Physics, School of Sciences, Ferdowsi University of Mashhad, Mashhad, 91775-1436, Iran}

\author{Shahram Abbassi}
\email{abbassi@ipm.ir}
\affiliation{Department of Physics, School of Sciences, Ferdowsi University of Mashhad, Mashhad, 91775-1436, Iran}
\affiliation{
School of Astronomy, Institute for Research in Fundamental Sciences (IPM), Tehran, 19395-5531, Iran}

\begin{abstract}
We investigate the properties of a hot accretion flow bathed in a poloidal magnetic field. We consider an axisymmetric viscous resistive flow in the steady state configuration. We assume the dominant mechanism of energy dissipation is due to turbulence viscosity and magnetic diffusivity. A certain fraction of that energy can be advected towards the central compact object. We employ self-similar method in the radial direction to find a system of ODEs with just one varible, $\theta$ in the spherical coordinates. 
For the existence and maintaining of a purely poloidal magnetic in a rotating thick disk, we find the necessary condition is a constant value of angular velocity along a magnetic field line. We obtain an analytical solution for the poloidal magnetic flux. We explore possible changes in the vertical structure of the disk under the influences of even symmetric and asymmetric magnetic fields. Our results reveal that a polar magnetic field with even symmetry about the equatorial plane makes the disk vertically thin. Moreover, the accretion rate decreases when we consider a strong magnetic field. Finally, we notice hot magnetized accretion flows can be fully advected even in a slim shape. 
\end{abstract}


\section{Introduction} \label{sec:intro}
Hot accretion flows are important to perceive hard X-ray spectrum generated around black holes, neutron stars and low luminous active galactic nuclei. In this kind of accreting system, radiation becomes inefficient to transport extra energy produced by turbulence out of the system. This feature makes it different from popular older model of Shakura \& Sunyaev's $\alpha-$disks. More potent differences between these two models are their vertical appearances, the first one is thought to be very thick ($H/R \sim$ 1) and the second one seems to be very thin ($H/R \sim $ 0.01).  The reason for the puffy shape of hot accretion flows is that the energy can't be radiated immediately after generation and it remains inside the disk and makes it warmer. There is another promising branch of accretion systems which is optically thick but geometrically thin which called slim diks (Abramowicz et al. 1988). The radiation plays an important role in these systems to transport energy outwards the disk beside being stored in the flow. Narayan \& Yi (1994) applied the energy equation as $Q^+-Q^-=fQ^+$ where, $Q^+, Q^-$ are heating, cooling energy rates, respectively, and $f$ is advection parameter which shows the fraction of energy advected towards the central compact object. It is common to write the hydrodynamics equations in cylindrical coordinates $(r,\phi,z)$ and use height-averaged quantities to remove dependency of $z$ from quantities. The radial structure of the disk is obtained by applying self-similarity technique. The more reliable way to study thick disks around black holes is to write basic equations in spherical coordinates which has been employed by Narayan \& Yi (1995a) (hereafter NY95).  Although NY95 presented the explicit structure of the fluid along the $\theta-$direction spherical coordinates, they didn't estimate the thickness of the disk. The flow in their work was assumed to fill the whole space between the two poles, hence the concept of disk and its thickness were not meaningful in that case. Gu et al. (2009) investigated the proper condition for advection dominance and found out the half-thickness of the disk must be greater than $72^\circ$ which results $H/R=tan(72^\circ)\sim 3$.  In the presence of magnetic field in the cylindrical coordinates, a  rough estimation of the disk's height is provided by presuming the hydrostatic balance in the vertical direction, that is, $H/R=(1+\beta) c_s/v_K$ where $\beta$ is the fraction of magnetic pressure to the gas pressure, $v_k$ is Keplerian velocity and $c_s$ is sound speed (e.g. Akizuki \& Fukue 2006, Abbassi et al. 2008, Ghasemnezhad et al. 2012, Ghasemnezhad 2017). If we include the magnetic force instead of the magnetic pressure, we can evaluate the disk's half-thickness more reasonable. Samadi, Abbassi \& Khajavi (2014) (hereafter SAK14) took into account a purely toroidal magnetic field and confirmed the thickness significantly reduced and the reason was magnetic pressure which acted in the opposite direction of gas pressure. 

The presence of magnetic fields in accreting systems is important because it is thought to be the source of viscous turbulence (via magneto-rotational instabilities, 'MRI') for moving angular momentum out of the flow (Balbus \& Hawley 1991, 1998). Furthermore, large-scale magnetic fields do not only controll the angular momentum transfer, but also drive mass outflow from the disks. Consequently, they can effectively change the structure of the accretion disks. The equipartition hypothesis for the magnetic energy density, proposed by Shvartsman (1971) showed that how large the field can become during the accretion process. Lovelace et al. (1986) developed a general theory for ideal MHD flows around black holes. They applied this theory for a thin steady, axisymmetric disk in cylindrical coordinates and later that approach was extended to study self-collimated jets from magnetized accretion disks with a/symmetric magnetic field (Lovelace et al. 1987, Wang et al. 1990, 1992). An analytical solution for Keplerian thin disks in the presence of a stellar magnetic field was obtained by Kaburaki (1986). In advection-dominated accretion flows, the temperature of the flow is much higher than thin disks and it makes this medium more capable to be affected by magnetic field. However, in the standard model of ADAFs (Narayan \& Yi 1995b), the only contribution of magnetic field is seen in the total pressure of the flow and its force and energy has been ignored. Afterwards, many athours considered magnetic force in momentum equation and employed the induction equation. Some of them (e.g. Shadmehri 2004,  Ghanbari et al. 2007, 2009, Shaghaghian 2011, 2016) studied purely poloidal magnetic fields and many concentrated only toroidal fields (Akizuki \& Fukue 2006, Khesali \& Faghei 2008, 2009, Mosallanezhad et al. 2014, 2016, Sarkar \& Das 2016) and a few theoritical works have been recently done with the global magnetic field (Samadi \& Abbassi 2016, Mosallanezhad et al. 2016, Samadi et al. 2017). 
% in general relativistic case e.g. Gammie et al. 2003, Camenzind 1986 

Regarding observations, the magnetic field is considerable in close binary systems for mass transfer and accretion (Zhilkin \& Bisikalo 2010). In X-ray binaries, the accretor is a neutron star with relatively weak magnetic field ($B\sim 10^4-10^6 G$ at the surface of it). Gosh and Lamb (1979a,b) were pioneers to study accretion onto magnetized neutron stars. To simplify the problem, they assumed a dipolar profile for the poloidal components of the magnetic field. They also used an ad hoc prescription for the toroidal magnetic field. After them, some authors (e.g. Wang 1987, Campbell 1987, 1992) investigated the proper estimation for the azimuthal component of $\textbf{B}$-field. Their result shows that the third component of magnetic field is generated in the axisymetric configuration, if only the rotational velocity changes in the vertical direction (Naso \& Miller 2010). In 2D simulation of Miller \& Stone (1997), it is found that the poloidal components of the magnetic field are very different from them in the dipolar one. Lubow et al. (1994) proposed a model for the evolution of a purely poloidal large-scale magnetic field. Okuzumi et al. 2014 derived the maximum strength of the steady poloidal field which was related to the disk size and imposed an upper limit on the accretion rate due to MRI (Takeuchi \& Okuzumi 2014).  

Magnetic field with poloidal configuration (stellar dipole or superposition of dipole and quadrupole) is very common for the simulations’ setup of accretion disks around young stellar objects (Dyda et al. 2015, Lovelace et al. 2010, Sheikhnezami et al. 2012, Goodson et al. 1999), massive young stars (Vaidya et al. 2011) and binary stars including a neutron star (Etienne et al. 2012, Wan 2017). Some other simulations whose main purpose is  jet launching adopt similar poloidal field geometry for hot accretion flows around black holes (e.g. Stone \& pringle 2001, Tchekhovskoy et al. 2011,  McKinney et al. 2012, Narayan et al. 2012,  Bu et al. 2016, Bai \& Stone 2013, Salvesen et al. 2016b).  A few works have been done with pure vertical magnetic fields (e. g. Stone \& Norman 1994, Beckwith et al.  2009, Bai \& Stone 2013, Suzuki T. K. \& Inutsuka 2014). In the most of these works, poloidal magnetic loops are twisted gradually due to differential rotation and consequently induces an azimuthal component for the magnetic field (Goodson et al. 1999). In some other simulations, the initial geometry of the magnetic field is assumed to be purely toroidal (e.g. Johansen \& Levin 2008 and Salvesen et al. 2016, Fragile \& ̨Sadowski 2017).  Salvesen et al. (2016) demonstrated that poloidal flux is a necessary initial condition for the sustainability of strongly magnetized accretion disks. Fragile \& ̨Sadowski (2017) obtain the same result which reveals that strongly magnetized toroidal fields are not sustainable and decay just after a short duration. General relativistic simulation is recently done by Gold et al. (2017) to investigate the magnetic field structure in Sgr A*. 

 Strictly speaking, the assumption of purely magnetic field is hardly possible because the differential rotation of the disk can easily make a toroidal field with shearing the radial component of the magnetic field, especially for ideal plasma with zero resistivity in steady state we can find $\nabla\times\textbf{B}=0\rightarrow \textbf{v}||\textbf{B}$ from the induction equation (Lubow et al. 1994). 
  However, one of our goals in this paper is finding proper conditions for this hardly expected configuration of $\textbf{B}$-field in a rotating thick disk with a finite conductivity.

In the present paper, we revisit the vertical structure of a hot accretion flow in the presence of a poloidal magnetic field. One of our purposes here is to determine the advection parameter with respect to polar angle. Narayan \& Yi 1994, 1995a,b assume that $f$ is constant for the whole of the flow whereas Gu et al. 2009 solve the ODE's set and then with using energy equation, they find angular profile of $f$. We follow the same way to obtain advection parameter versus polar angle. In the next section, we derive basic equations. We simplify the partial differential equations with self-similar technique in section 3. For solving ODEs we introduce proper boundary conditions in section 4. The vertical structure of the disk is presented in section 5. A brief summary and conclusion come in section 6. 

\section{Equations}
In order to study the possible effect of  poloidal magnetic field on the accretion system, we follow Gu et al. 2009 and assume $v_\theta=0$ which provides more accuracy in the spherical coordinates in comparison with accuracy of considering $v_z=0$ in the cylindrical coordinates. On the other hand, in the spherical coordinate we do not need to use height-averaged quantities which is less reliable in thick disks. Other assumptions in our calculation are axisymmetry $\partial/\partial \phi=0$, steady state $\partial/\partial t=0$ and $\Phi=-GM_*/r$ which is the gravitational potential of just the central object. Moreover, we neglect the relativistic effects. Now, we can write the basic equations for these quantities, density: $\rho$, gas pressure: $p$, velocity vector: $\textbf{v}=v_r \hat{e}_r+v_\phi \hat{e}_\phi$, poloidal magnetic field:  $\textbf{B}=B_r \hat{e}_r+B_\theta \hat{e}_\theta$. The first equation is related to mass conservation $\partial \rho/\partial t+\nabla\cdot(\rho\textbf{v})=0$ which is simplified as below with our mentioned assumptions, 
\begin{equation}
\frac{1}{r^2}\frac{\partial}{\partial r}(r^2 \rho v_r)=0,
\end{equation}
The momentum equation,$\rho D\textbf{V}/Dt=-\nabla p - \rho\nabla\Phi-\nabla\cdot\textbf{T}^\nu+(\textbf{J}\times \textbf{B})/c$ ($\textbf{T}^\nu$ is stress tensor) yields three scalar equations,
    \begin{equation}
    v_r\frac{\partial v_r}{\partial r}-\frac{v_\phi^2}{r}
  =-\frac{GM_*}{r^2}-\frac{1}{\rho}\frac{\partial p}{\partial r}+\frac{1}{c\rho}(J_{\theta}B_\phi-J_{\phi}B_\theta)
\end{equation}
\begin{equation}
    -\frac{v_\phi^2}{r} \cot\theta
   =-\frac{1}{\rho r }\frac{\partial p}{\partial \theta}+\frac{1}{c\rho }(J_\phi B_r-J_r B_\phi)
\end{equation}
\begin{equation}
\frac{v_r}{r}\frac{\partial(r v_\phi)}{\partial r}
 =\frac{1}{\rho r^3}\frac{\partial}{\partial r}(r^3 T_{r\phi})+\frac{1}{c\rho}(J_r B_\theta-J_\theta B_r),
\end{equation}
We presumed just $r\phi$ component of the stress tensor, $T_{r\phi}$ is important and under $\alpha$- prescription (Shakura \& Sunyaev 1973), it is defined by 
\begin{displaymath}
T_{r\phi}=\nu\rho r \frac{\partial}{\partial r}(\frac{v_\phi}{r}),\hspace{0.3cm}\nu=\alpha c_s H
\end{displaymath}
where $\nu$ is the kinematic viscosity coefficient, $\alpha$ is the constant viscosity parameter, $c_s$ is sound speed and $H=r c_s/v_K$ is a typical value of the height scale in an accretion disk, $v_K =\sqrt{GM/r}$ is the Keplerian velocity.
\begin{equation}
T_{r\phi}=\frac{\alpha r^2}{v_K} \frac{\partial}{\partial r}(\frac{v_\phi}{r})\rho c_s^2 
\end{equation}
Now we need to consider Maxwell equations, the first one is the induction equation: $\partial\textbf{B}/\partial t=\nabla\times(\textbf{v}\times\textbf{B}-\eta\nabla\times\textbf{B})$, which can be simplified with our assumptions as,
\begin{equation}
v_rB_\theta=\frac{\eta}{r }\bigg[\frac{\partial(rB_\theta)}{\partial r}-\frac{\partial B_r}{\partial\theta}\bigg]
\end{equation}
the above equation is the result of $\partial B_r/\partial t=0$ and $\partial B_\theta/\partial t=0$  (for more details refer to Samadi \& Abbassi 2016). To satisfy $\nabla\cdot\textbf{B}=0$ we can express the poloidal magnetic field in terms of a flux function $\psi(r,\theta)$ (Lovelace et al. 1987), 
\begin{displaymath}
\textbf{B}_p=\frac{1}{r\sin\theta}\nabla\psi\times\hat{e}_\phi
\end{displaymath}
A little manipulation leads us to have two seperate equations for $B_r$ and $B_\theta$ as functions of $\psi$,  
\begin{equation}
B_r=\frac{1}{r^2\sin\theta}\frac{\partial\psi}{\partial\theta}, \hspace{0.5cm}
B_\theta=-\frac{1}{r\sin\theta}\frac{\partial\psi}{\partial r}
\end{equation}
Notice a constant value of $\psi$ can form a 3D surface by rotating a magnetic field line around polar axis.
To find the current density vector, we need another relation of Maxwell's equations, $\textbf{J}=c/4\pi (\nabla\times\textbf{B})$. With a purely poloidal magnetic field, we can see $\textbf{J}$ has just one non-zero component,
\begin{equation}
J_r=J_\theta=0, \hspace{0.5cm}J_\phi=\frac{c}{4\pi r}\bigg[\frac{\partial}{\partial r}(rB_\theta)-\frac{\partial B_r}{\partial\theta}\bigg]
\end{equation} 
 Finally we ought to know the magnetic diffusivity, $\eta$. It is thought that turbulence can be the origin of magnetic resistivity, hence we can write a relationship for $\eta$ the same as $\alpha$- prescription of Shakura \& Sunyaev (1973) for the turbulent viscosity (Shadmehri 2004, Bisnovatyi-Kogan \& Ruzmaikin 1976),
\begin{equation}
\eta=\frac{\eta_0 r}{v_K} c_s^2 
\end{equation}
In the next section we will use self-similarity in the radial direction to eliminate $r-$dependency from our quantities, then we will obtain a system of ordinary differential equations with respect to one independent variable of $\theta$.
 \begin{figure*}
\centering
\includegraphics[width=155mm]{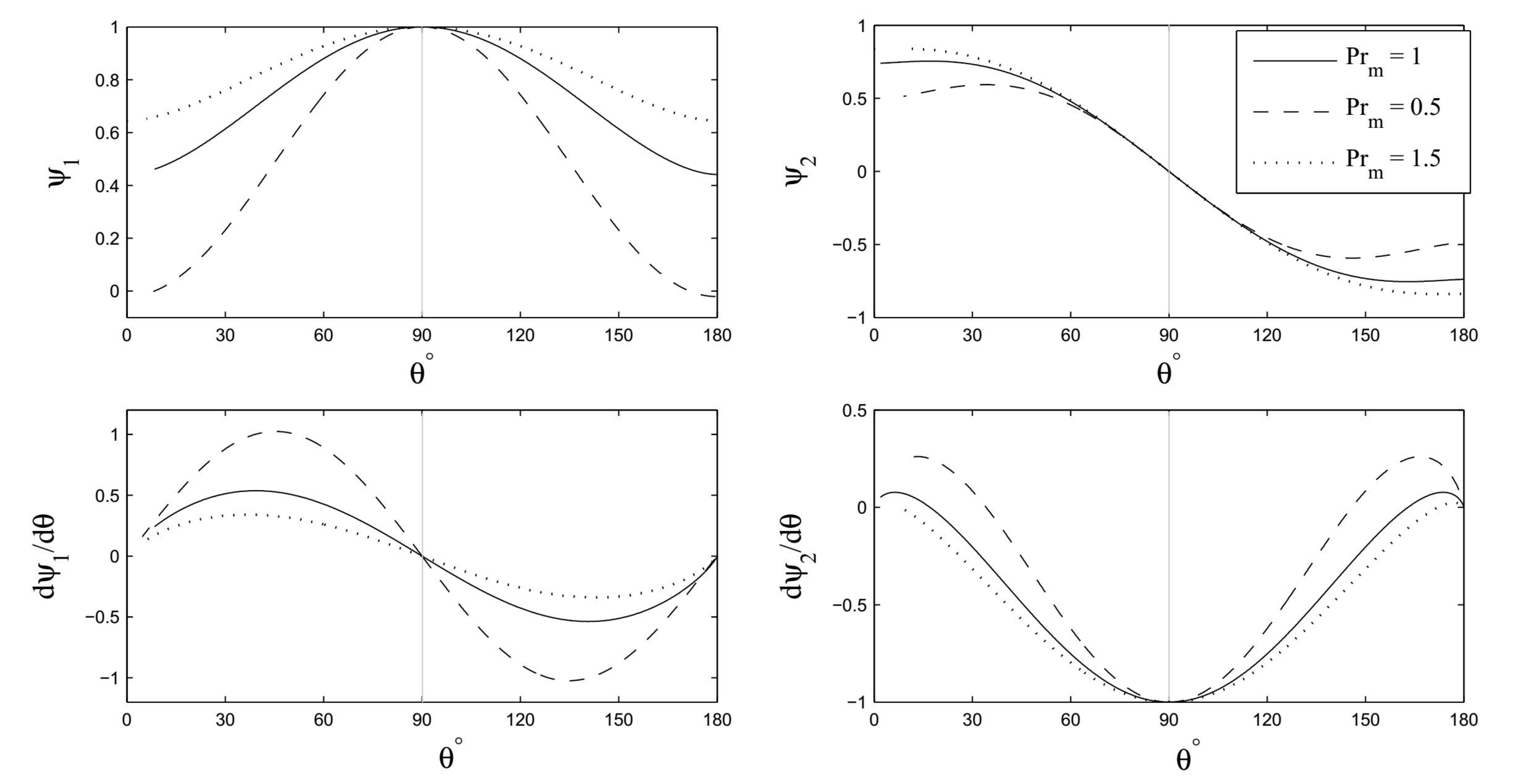}
 \caption{The two solutions of Eq. (14); $\psi_1$ has an even symmetry about the equatorial plane whereas $\psi_2$ has an odd symmetry. We have shown the magnetic flux function $\psi$ and its derivative $d\psi/d\theta$ with 3 values of Prandtl number.  }
\end{figure*}

\section{Self-Similar Solutions}
Following  NY95a and the other similar works (Ghanbari et al. 2007, 2009, SAK14),  self-similarity in the radial direction is employed, so all types of velocities become proportional to $r^{-1/2}$ and for density we can write, $\rho \propto r^{-n}$. If we use $v_r(r,\theta)=v_r(\theta) r^{-1/2}$ and $\rho(r,\theta)=\rho(\theta) r^{-n}$ in the continuity equation (1), we will have:
\begin{displaymath}
\rho(\theta) v_r(\theta)\frac{d}{dr}(r^{2-n-1/2})=0\rightarrow n=\frac{3}{2}
\end{displaymath}
According to the formula of $p\propto\rho c_s^2$, gas and magnetic pressure must be proportional of radius as $(p,B_i^2) \propto r^{-3/2-1}$, then with using $p_m=B^2/8\pi$ we can find $B_i\propto r^{-5/4}$ and $\psi\propto r^{3/4}$. By substituting Eq. (6)-(8) and considering the self-similar dependency of every quantity, Eq. (2)-(4) are reduced to be,
  \begin{equation}
\frac{1}{2}v^2_r+v^2_\phi
 - v^2_k+\frac{5}{2}\frac{p}{\rho} =\frac{rv_r}{4\pi\rho\eta}B_\theta^2 ,
\end{equation}
 \begin{equation}
v^2_\phi \cot\theta 
  -\frac{1}{\rho}\frac{dp}{d\theta}
+\frac{rv_r}{4\pi \rho\eta}(B_r B_\theta)=0,
 \end{equation}
\begin{equation}
v_r =-\frac{3\alpha}{2v_k} c_s^2
\end{equation}
The relations of Eq. (7) with self-simlar solutions become,
\begin{equation}
B_r=\frac{1}{r^2\sin\theta}\frac{d\psi}{d\theta}, \hspace{0.5cm}
B_\theta=-\frac{3\psi}{4r^2\sin\theta}
\end{equation}
The third component of the induction equation has still remained, $\partial B_\phi/\partial t=0$, which is simplified as:
\begin{displaymath}
\frac{d}{d\theta}( B_\theta v_\phi )-\frac{3}{4}B_rv_\phi =0,
\end{displaymath}
With using relations of (13), we obtain,
\begin{displaymath}
\frac{d}{d\theta}(\frac{v_\phi}{r\sin\theta}\psi)+\frac{v_\phi}{r\sin\theta}\frac{d\psi}{d\theta}=0 ,
\end{displaymath}
Multiplying the above equation by $\psi$, we find a new constant parameter, $K$, as
\begin{equation}
\frac{d}{d\theta}(\frac{v_\phi}{r\sin\theta}\psi^2)=0 \rightarrow \frac{v_\phi}{r\sin\theta}\psi^2=K
\end{equation}
\begin{figure*}
\centering
\includegraphics[width=165mm]{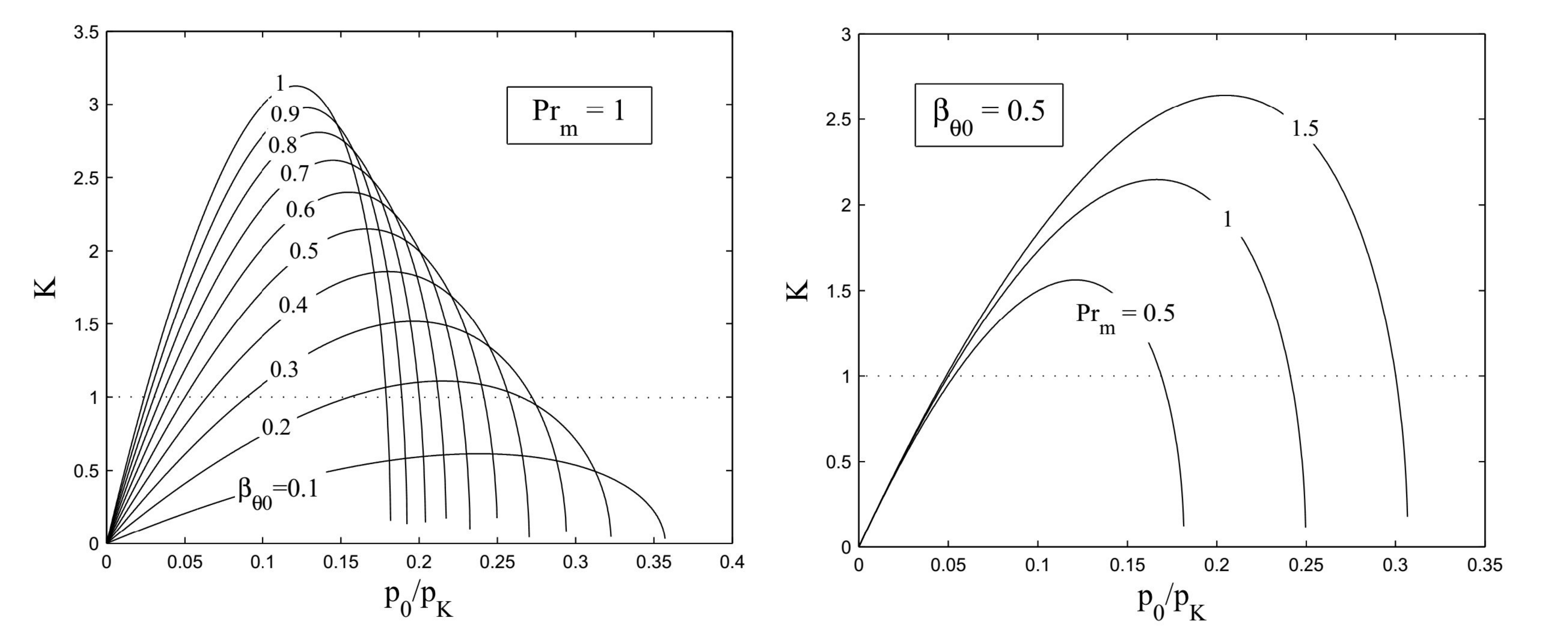}
 \caption{Variation of the parameter of $K$ ($=\psi_0 ^2 v_{\phi0}$) with the gas pressure in the equatorial plane $p_0/p_K$ under the influence of different a) poloidal magnetic fields on the left side panel and b) Prandtl number on the right side panel. In this figure, $p_K=\rho_0 v_K^2$ is the fiducial pressure, $v_K$ is the Keplerian velocity and $\beta_{\theta0}=B_{\theta0}^2/8\pi p_0$. The other input parameters are $\gamma=1.5, \alpha=0.1$ and $\beta_{r0}=0$. It is clear that there is a maximum value for $K$.}
\end{figure*}
 The above relationship between $\psi$ and $v_\phi$ says, the flow rotates faster where the magnetic field becomes weaker. It is interesting to point out that Lovelace et al. (1986) acheived similar dependency between these two different quantities. They found out the angular rotation of the disk matter is a function of the magnetic flux, $\psi$. Moreover, we see the same relationship $\Omega=f(\psi)$ (Eq. 39) in the work of Ferraro (1937) which means that the rotational velocity is constant on a field line in a steady state. Here, we return to equation (6), 
\begin{equation}
\frac{d^2\psi}{d\theta^2}=\cot\theta\frac{d\psi}{d\theta}+\frac{3}{4}(\frac{rv_r}{\eta}+\frac{1}{4})\psi
\end{equation}
Considring Eq. (9), (12) we have, $rv_r/\eta=-3/(2Pr_m)$, where $Pr_m=\eta_0/\alpha$ is Prandtl number. Substitution into Eq. (15) yields  the second order differential equation with only one unknown, i.e. $\psi$,
\begin{equation}
\frac{d^2\psi}{d\theta^2}=\cot\theta\frac{d\psi}{d\theta}+\frac{3}{4}(-\frac{3}{2 Pr_m}+\frac{1}{4})\psi
\end{equation}
This ODE has an analytic solution in the form of hypegeometric functions, 
\begin{equation}
\psi_1(\theta)=\sin\theta^2 \hspace{0.05cm} _2F_1\bigg(\frac{3}{4}+\frac{\epsilon}{8}, \frac{3}{4}-\frac{\epsilon}{8}; \frac{1}{2};\cos^2\theta\bigg)
\end{equation}
\begin{equation}
\psi_2(\theta)=\sin^2\theta \cos\theta\hspace{0.05cm} _2F_1\bigg(\frac{5}{4}+\frac{\epsilon}{8}, \frac{5}{4}-\frac{\epsilon}{8}; \frac{3}{2};\cos^2\theta\bigg)
\end{equation}
where $\epsilon=\sqrt{18 /Pr_m+1}$. The similar result was obtained by Okuzumi et al. (2014) for the disk-induced magnetic flux.

Figure 1 shows the two solutions of Eq.(14), $\psi_1$ and $\psi_2$ and their derivatives. These two magnetic flux functions are symmetric about the equator. $\psi_1$ is even and $\psi_2$ is odd. The general solution of Eq.(14) is a composition of these two possible solutions, that is
\begin{equation}
\psi=C_1\psi_1+C_2\psi_2
\end{equation}
%For a general case, we will consider a composition of $\psi_1$ and $\psi_2$, i.e. $\psi=C_1 \psi_1+C_2 \psi_2$.
 One more step is required to solve the differential equation (11) to determine the gas pressure. Velocities are known from Eq. (12), (14) and magnetic field components can be substituted by Eq. (17), (18) and using Eq.(7). 
Finally, the differential equation (11) can be numerically solved with proper boundary conditions as seen in the following section.

\section{Boundary conditions}
In this section, we focus on the initial values of $\rho, p, v_r, v_\phi$ in a certain angle of $\theta=90^\circ$ which specifies the equator. It is common to use reflection symmetry about the equatorial plane. However, it can be possible just in the presence of a symmetric magnetic field. In this work, we firstly suppose a poloidal magnetic field and secondly we assume a general case of magnetic field with asymmetric configuration.  
\subsection{Symmetric Poloidal Magnetic Field}
In the presence of a poloidal magnetic field, the magnetic flux $\psi$ has even symmetry configuration. For this kind of field, we should use $\psi_1$ and neglect $\psi_2$ or select $C_2=0$,
\begin{equation}
\psi=C_1\psi_1
\end{equation}
 Moreover, the total magnetic field at the equator becomes in the meridional direction, so
\begin{equation}
B_{r0}=0,\hspace*{0.3cm} B_{\theta0}=\sqrt{8\pi\beta_{\theta0} p_0}
\end{equation}
where $\beta_{\theta0}$ is a parameter that shows the ratio of magnetic pressure ($p_B=B^2/8\pi$) to gas pressure at the equatorial plane (zero index means the value of quantity in $\theta=90^\circ$). 
 Consequently, from Eq.(13), (20) and (21) we obtain $C_1$ with respect to $\beta_{\theta0}$,
 \begin{equation}
C_1 =-\frac{4 r^2}{3}\sqrt{8\pi\beta_{\theta0} p_0}
\end{equation} 
therefore the components of this magnetic field from Eq. (13) become,
\begin{displaymath}
\end{displaymath} 
 \begin{equation}
 B_r=-\frac{4 \sqrt{8\pi\beta_{\theta0} p_0}}{3\sin\theta}\frac{d\psi_1}{d\theta},\hspace{0.5cm}
B_\theta=\sqrt{8\pi\beta_{\theta0} p_0}\frac{\psi_1}{\sin\theta}
 \end{equation}
From Eq. (11) we can find the derivative of gas pressure at the boundary,
\begin{equation}
\frac{dp}{d\theta}|_{_{\theta=90^\circ}}=0,
\end{equation}
Eq. (12) leads us to obtain $v_{r0}$ as,
\begin{equation}
v_{r0}=-\frac{3\alpha}{2v_k} c_{s0}^2,
\end{equation} 
Magnetic diffusivity in the equator from Eq. (9) becomes,
\begin{equation}
\eta|_0=\frac{\eta_0 r}{v_K}c_{s0}^2,
\end{equation}
 To obtain $v_{\phi0}$ we can use Eq. (14) and (19)
\begin{equation}
v_{\phi0}=\frac{rK}{C_1^2\psi_{10}^2}=\frac{9K}{128\pi r^3 \beta_{\theta0} p_0}
\end{equation}
\begin{figure*}
\centering
\includegraphics[width=165mm]{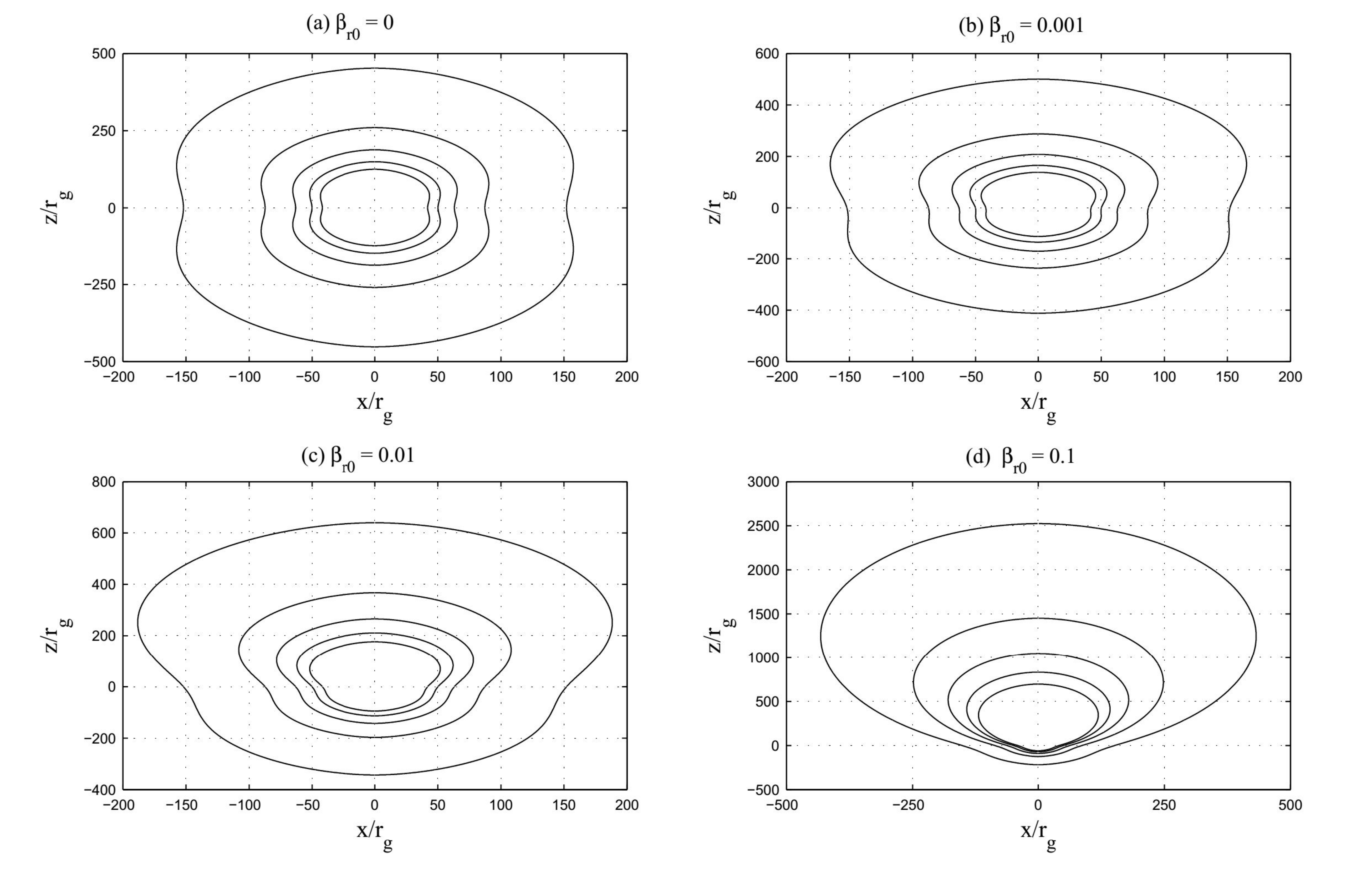}
 \caption{The presentation of magnetic field lines for both symmetric and asymmetric configurations about the equatorial plane, (the horizontal axis is located at the equator). (a) symmetric magnetic field corresponding to $\beta_{r0}=0$, (b-c) asymmetric magnetic field lines. The other input parameters are  $\beta_{\theta0}=0.3$, $\gamma=1.5$, $Pr_m=1$, $\alpha=0.1$ and $K=1$. }
\end{figure*} 

 Now we can substitute relations of (23), (25) and (27) in Eq. (10) to find $K$,
\begin{equation}
K=\frac{128\pi r^3}{9}\beta_{\theta0}p_0\sqrt{v_K^2-\frac{9\alpha^2}{8}\frac{c_{s0}^4}{v_K^2}-(\frac{5}{2} +\frac{3\beta_{\theta0}}{Pr_m}) \frac{p_0}{\rho_0}}
\end{equation}
where $c_{s0}=\sqrt{\gamma p_0/\rho_0}$ is the sound speed at the equator. In the above relation, $p_0$ and $\rho_0$ are still unknown but we set $\rho_0=1$ and this implies a characteristic scale density at $\theta=90^\circ$ (Xue \& Wang 2005). To determine the possible values of $p_0$ we follow the same way as one in SAK14. Consequently, taking into account the minimum possible of $v_{\phi0}^2$ and $v_{r0}^2$ in Eq. (10) helps us to guesstimate the maximum possible value of gas pressure at the boundary, that is $p_0=0.4\rho_0 v_K^2$. In figure 2, we have plotted $K$ with respect to $p_0/p_K$, where $p_K=\rho_0 v_K^2$. We see how much $K$ varies due to firstly magnetic field and secondly Prandtl numbers. This figure shows for a certain poloidal magnetic field with an even symmetry configuration there is a maximum value of $K$ which can be less than unique in the presence of weak fields with $\beta_{\theta0}<0.2$.  Furthermore, we can see the permissible range of $p_0$ has become smaller in the presence of the stronger magnetic field and we will see (in Fig. 10) this means the flow's thickness shrinks when $\beta_{\theta0}$ is bigger. According to the right side panel of this figure, the maximum value of $K$ increases when Prandtl number is larger. 

\subsection{Asymmetric Poloidal Magnetic Field}
The general situation of a 2D magnetic field is a combination of even and odd symmetric configurations,
\begin{displaymath}
\psi=C_1\psi_1+C_2\psi_2
\end{displaymath}
From Fig.1 we know 
\begin{displaymath}
\psi_1(\theta=90^\circ)=1, \hspace{0.5cm}\psi_2(\theta=90^\circ)=0,
\end{displaymath}
\begin{displaymath}
\frac{d\psi_1}{d\theta}|_{\theta=90}=0,\hspace{0.5cm}\frac{d\psi_2}{d\theta}|_{\theta=90}=1
\end{displaymath}
Then we use Eq.(13) to find the components of the magnetic field at the equatorial plane,
\begin{displaymath}
B_{r0}=\frac{C_2}{r^2},\hspace{0.5cm} B_{\theta0}=-\frac{3C_1}{4r^2}
\end{displaymath}
 On the other hand, from the magnetic pressure we can define the magnetic components with respect to the gas pressure and the ratio of the magnetic pressure, $\beta_{r0}=B_{r0}^2/(8\pi p_0)$, $\beta_{\theta0}=B_{\theta0}^2/(8\pi p_0)$, hence we obtain the constants in the magnetic flux,
 \begin{equation}
 C_1=-\frac{4r^2}{3}\sqrt{8\pi\beta_{\theta0}p_0},\hspace{0.5cm} C_2=r^2\sqrt{8\pi\beta_{r0}p_0}
 \end{equation}
 Figure 3 displays the magnetic field lines for (a) even symmetric poloidal magnetic field  and (b)-(d) asymmetric ones. To draw the magnetic field lines, we have used the relation of $dr/B_r=rd\theta/B_\theta$ (Abbassi \& Ghanbari 2004). Although we have shown magnetic field lines in the whole space between the equator and the disk's poles, there is an issue and that is the disk's opening angle (where the gas pressure becomes zero) is not zero (nor $\pi$), hence Eq. (12) and (16) are just valid inside the disk.      

\begin{figure*}
\centering
\includegraphics[width=165mm]{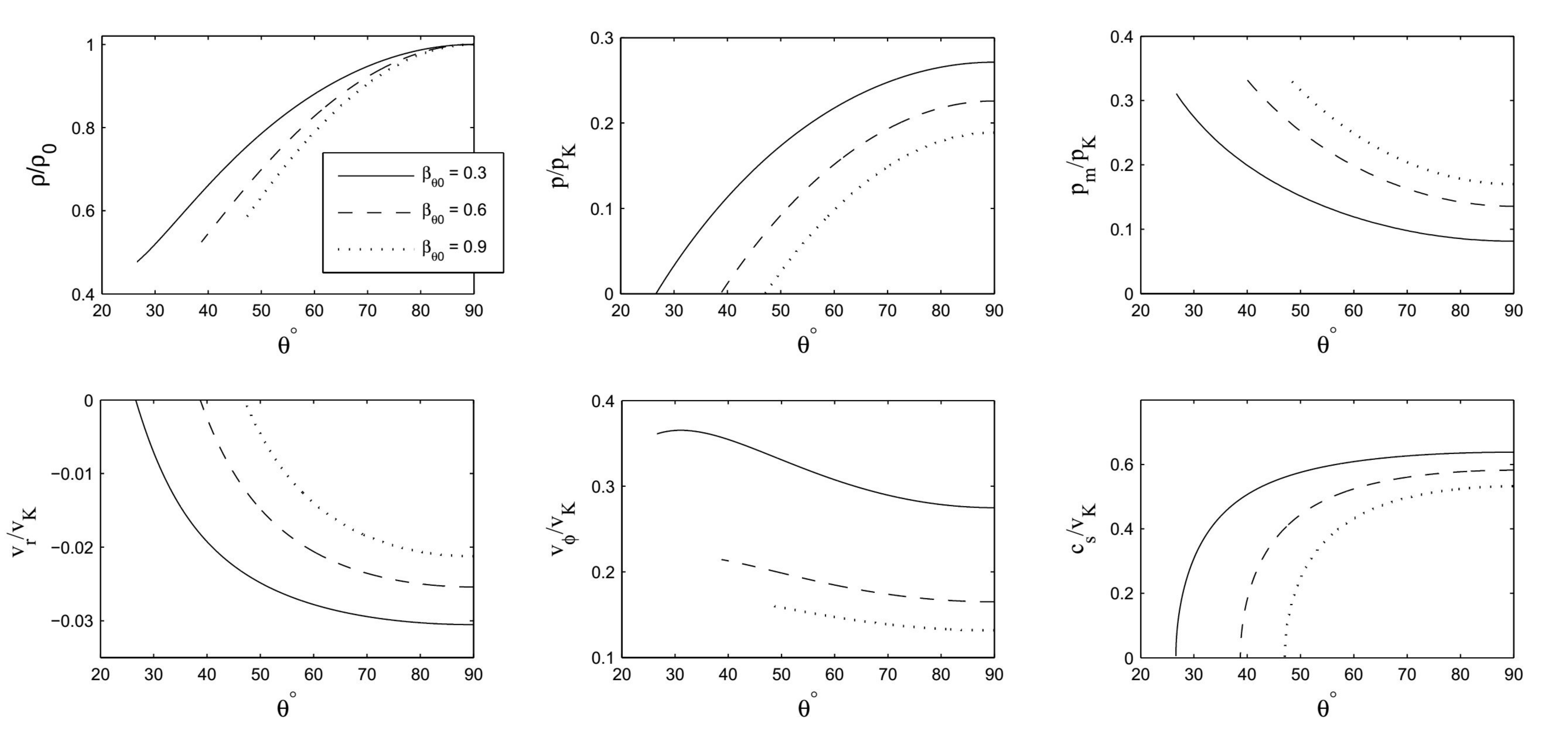}
 \caption{Self-similar solutions (corresponding to  $Pr_m=1, \alpha=0.1, \gamma=1.5$ and $K=1$) in the presence of three symmetric poloidal magnetic fields with $\beta_{r0}=0$. In this figure, the disk's surface is located in a certain angle which the gas pressure becomes zero. Here, $v_K$ is Keplerian velocity and $p_K$ is  the fiducial pressure and determined by $p_K=\rho_0 v_K^2$.}
\end{figure*}
\begin{figure*}
\centering
\includegraphics[width=140mm]{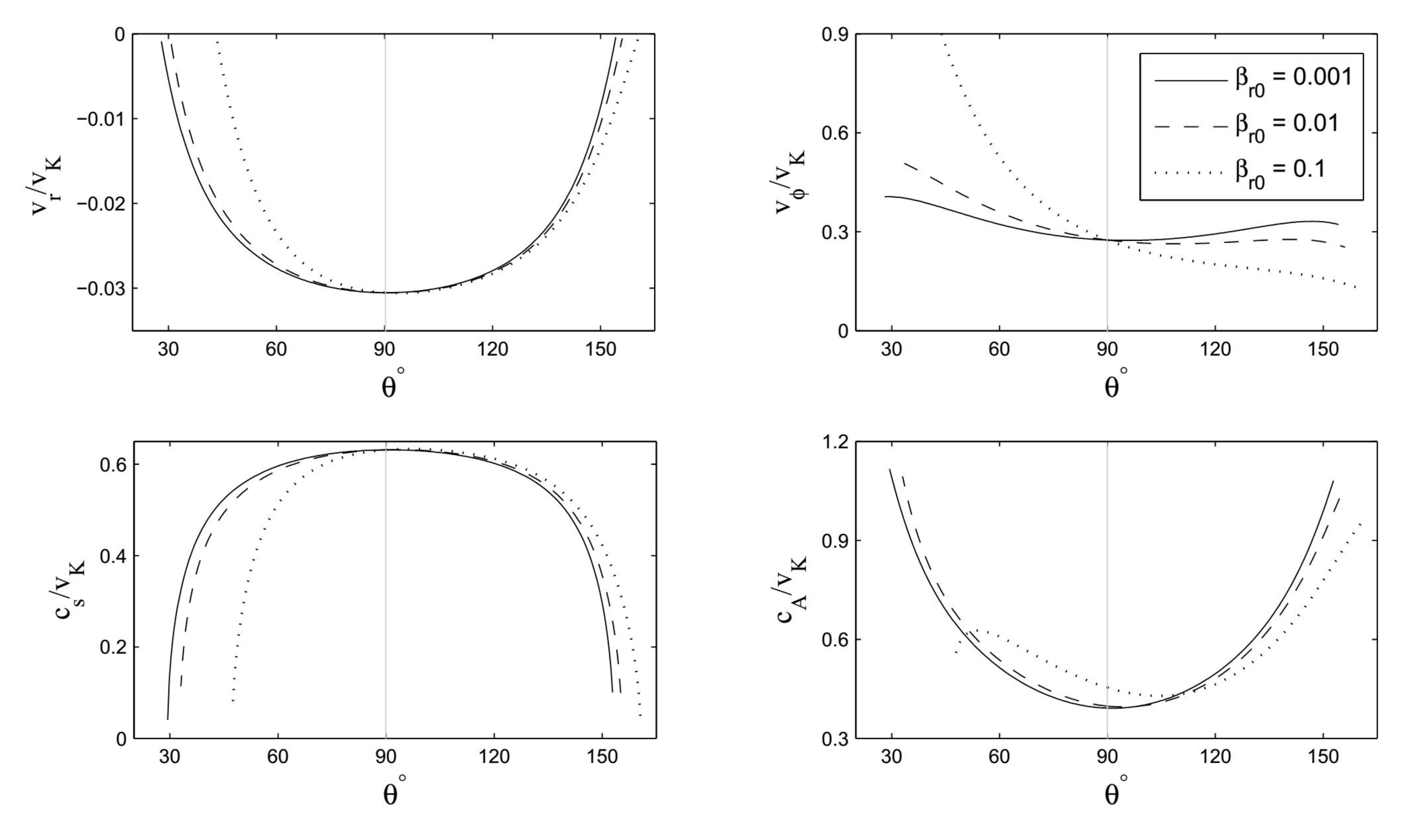}
 \caption{Self-similar solutions (corresponding to  $Pr_m=1, \alpha=0.1, \gamma=1.5$ and $K=1$) in the presence of three asymmetric magnetic field with $\beta_{r0}\neq 0$. Here, $c_A$ is Afven velocity which is defined by $\sqrt{(B_r^2+B_\theta^2)/(4\pi\rho)}$. }
\end{figure*}

\begin{figure*}
\centering
\includegraphics[width=140mm]{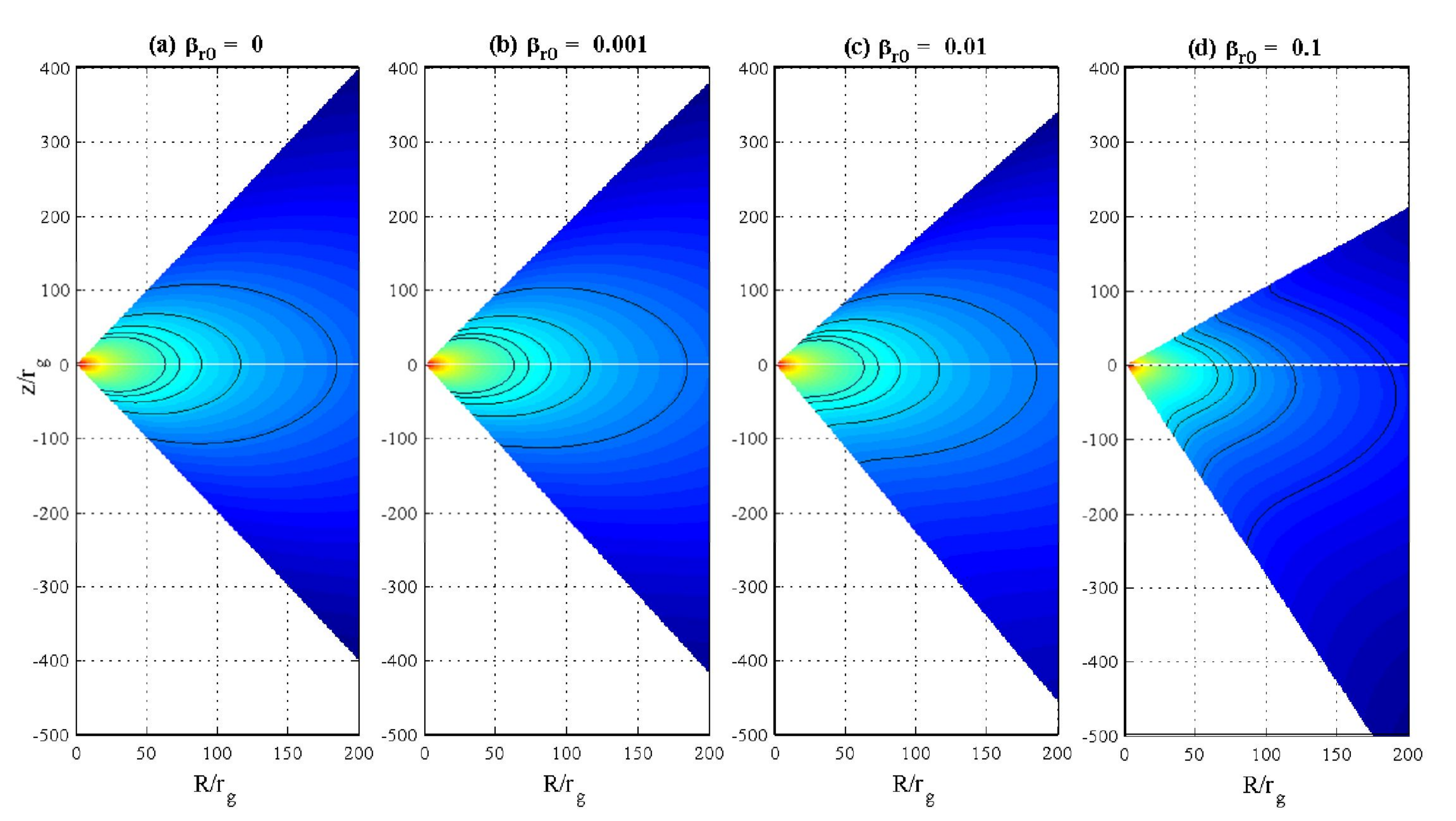}
 \caption{Density contours in the meridional plane in the presence of asymmetric magnetic fields. Here, the input parameters are  $Pr_m=1, \alpha=0.1, \gamma=1.5, K=1$ and $\beta_{\theta0}=0.3$. The difference in half-thickness of disk above and below the equatorial plane ($z=0$)  is clear especially in the right panel.}
\end{figure*}
\begin{figure*}
\centering
\includegraphics[width=140mm]{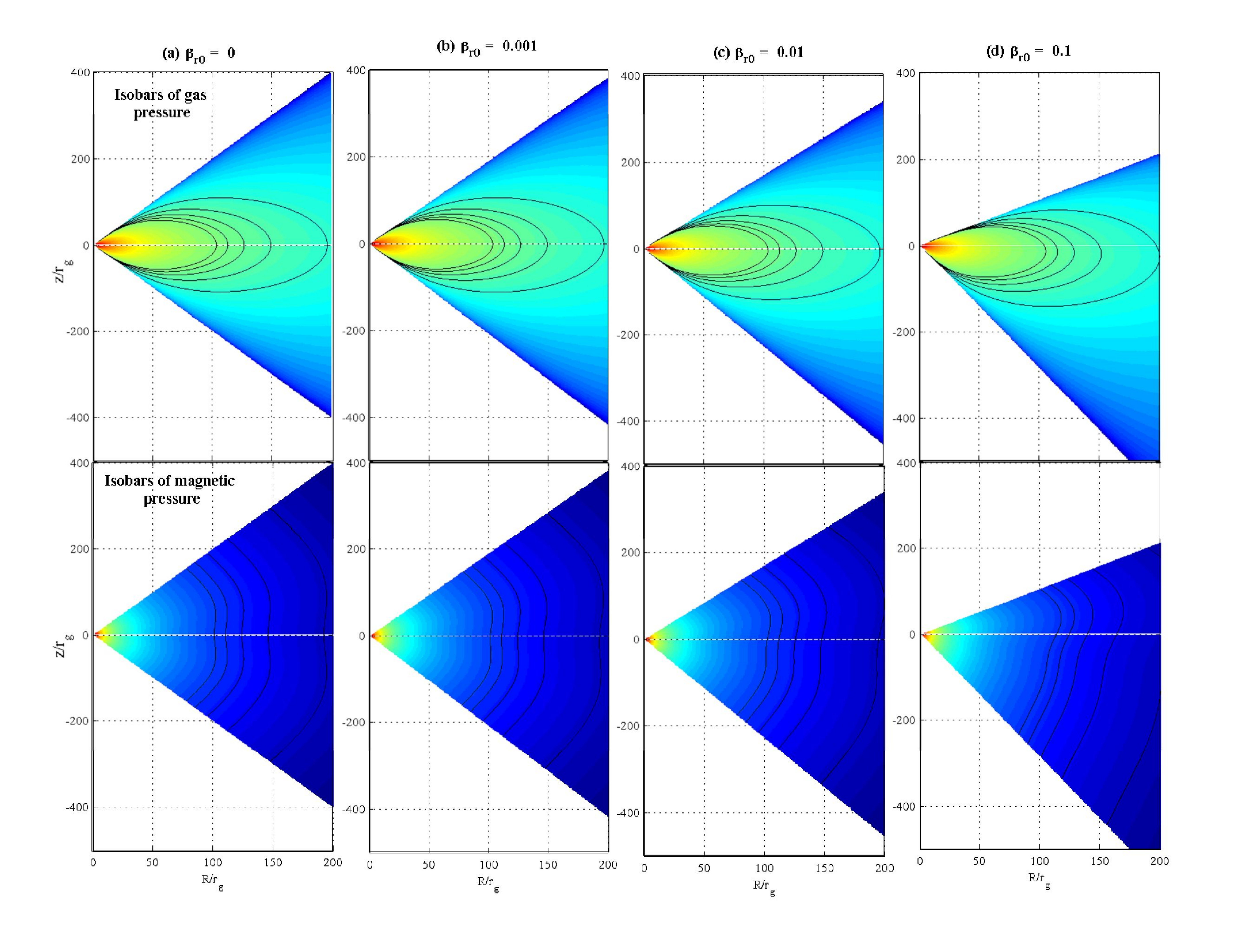}
 \caption{Isobar surfaces of gas pressure (top) and magnetic pressure (bottom) in 2D space, for a typical solution with parameters $Pr_m=1, \alpha=0.1, \gamma=1.5$ and $K=1$. The asymmetric magnetic fields are provided with $\beta_{\theta0}=0.3$ and non zero values of $\beta_{r0}$. }
\end{figure*}
\begin{figure*}
\centering
\includegraphics[width=140mm]{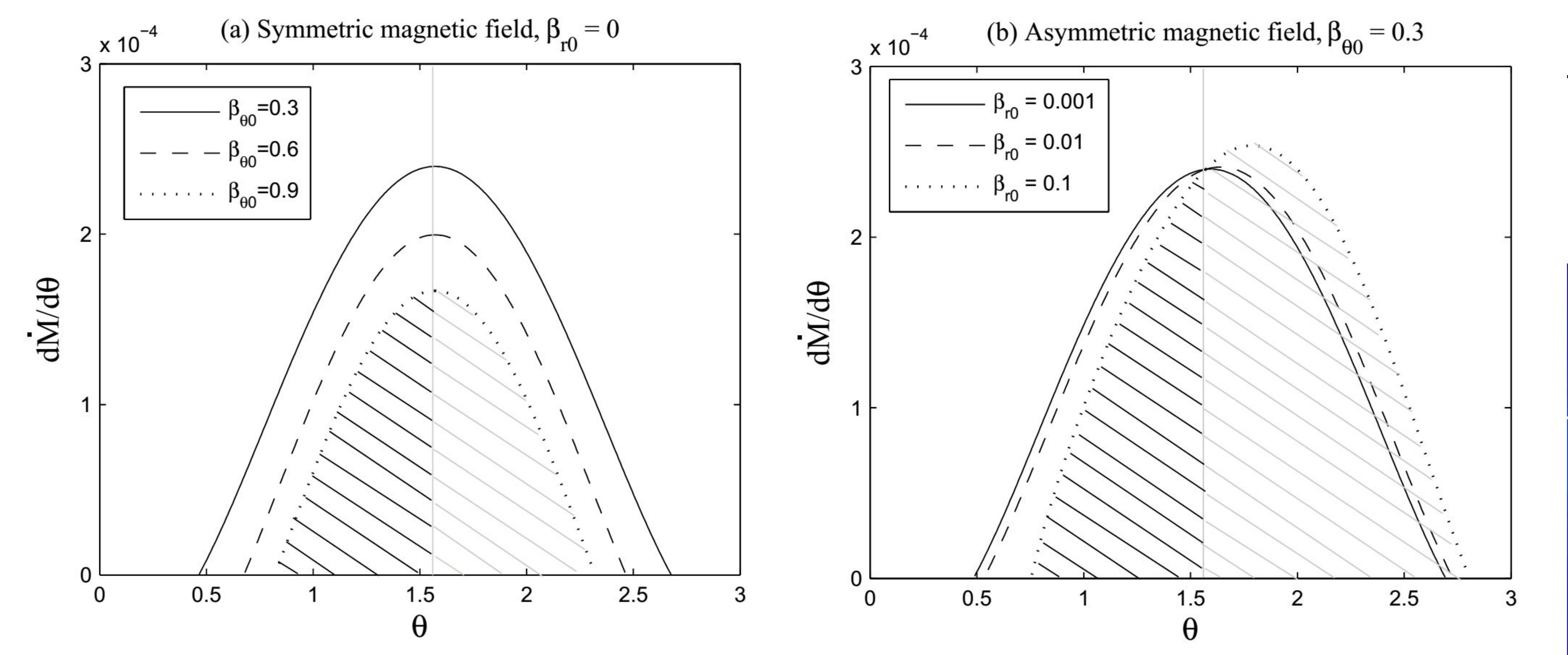}
\caption{The distribution of mass accretion rate differentiation along $\theta-$direction corresponding to $\gamma=1.5$, $\alpha=0.1, Pr_m=1, K=1$. \textbf{Left side:} The effect of even symmetric magnetic fields with $\beta_{r0}=0$;  The area between the horizontal axis and the curve shows the total accretion rate, we have shown it with shaded area for the case of $\beta_{\theta0}=0.9$. \textbf{Right side:} The asymmetric magnetic fields effect with $\beta_{\theta0}=0.3$;  We have shown the total accretion rate for the case of $\beta_{r0}=0.1$ with shaded area. }
\end{figure*}

Comparison with symmetric case, just one difference in other boundary conditions appears from Eq. (12). When we write it in $\theta=90^\circ$, the first term in the left side becomes zero, then we will have,
\begin{equation}
 \frac{dp}{d\theta}|_0=
\frac{3\alpha}{8\pi\eta_0}B_{r0} B_{\theta0}=\frac{3\sqrt{\beta_{r0}\beta_{\theta0}}}{Pr_m}p_0,
 \end{equation}
\section{Vertical structure}
Using proper boundary values of the physical quantities from the previous section, we can integrate numerically the equation of (11) with respect to polar angle $\theta$ to obtain $p$. We stop the integration where the gas pressure becomes zero. On the other hand, $\rho, v_r$ and $v_\phi$ can be calculated from Eq. (10), (12) and (14). Fig. 4 shows how scalar quantities vary in the meridional direction inside the disk and also how much a stronger magnetic field affect them. As we expect, the density and gas pressure reduces towards the surface of the disk but magnetic pressure is minimum at the equator and becomes larger and larger near the surfaces (for both cases of symmetric and asymmetric fields). These opposite trends of gas and magnetic pressures are also reported in the presence of purely toroidal magnetic field (SAK14). In the second row of Fig.4, the two components of the flow's velocity, $v_r$ and $v_\phi$ are minimum at the equator and increase towards the surface. Moreover, the kinetic energy decreases in the larger symmetric $\textbf{B}-$field because both $v_r$ and $v_\phi$ have smaller magnitudes with a larger $\beta_{\theta0}$. The same result had been obtained in the presence of a purely toroidal magnetic field in SAK14 but the only difference was the location of disk's surface which was in a certain angle with zero value of rotational velocity. Other important quantity is sound velocity whose squared magnitude defines the flow's temperature. Generally, the sound velocity decreases from the mid-plane to outwards. The effect of even symmetric field is significant on $c_s$ and it is seen that sound speed has  shifted to lower values in the presence of the stronger field. Therefore, the disk becomes colder with a larger $\beta_{\theta0}$. 
The general feature of this figure is changing the half-thickness of the disk due to different values of $\beta_{\theta0}$. In fact, the presence of a stronger magnetic field makes the disk thinner and limits the gas pressure value in the mid-plane.    
 
In the first row of figure 5, the radial and rotational components of velocity  are presented by taking into account  asymmetric magnetic fields. In asymmetric configuration of magnetic field, it is seen that the radial velocity has still the same ascending trend towards the surface but it doesn't have reflection symmetry about $\theta=90^\circ$. On the contrary, $v_\phi$ has opposite behavior in northern and southern hemispheres of the disk. In the second row of this figure, $c_s$ and $c_A$ behave like $p$ and $p_m$, respectively. Similar to symmetric $\textbf{B}-$field, the sound speed is minimum at $\theta=90^\circ$, nevertheless Alfven velocity comes up very different in one side of the disk when $\beta_{r0}$ is rather large.  

Figure 6 displays isodensity contours for the asymmetric configuration of disk which is caused by asymmetric poloidal magnetic field. In the reflection symmetry of quantities, isodensity contours usually make close shapes such as Fig.2 of Samadi et al. (2016). But here, non-zero $B_r$ in the mid-plane makes isodensity contour to be different in upper and lower sides of the disk. In the case of $\beta_{r0}=0.1$, the shape of isodensity contours are not close, that means the density unusually becomes larger at the bottom surface. The different half-thicknesses (or height) of the disk is seen better here. 

The isobars of gas and magnetic pressures are shown in figure 7. The close shapes of gas pressure contours imply that the general behavior of this quantity doesn't change significantly in the presence of an asymmetric magnetic field, although the reflection symmetry about equatorial plane gradually disappears with large $\beta_{r0}$. In the second row of Fig. 7, the isobar contours of magnetic pressure have completely different shape in comparison with isobars of gas pressure and that is because of opposite trends of these two quantities even in non-symmetric configuration of magnetic field. In many cases, the magnetic pressure at the surface is larger than the disk's equator.
 
\subsection{Mass Accretion Rate}
As we know, the luminosity of the accretion disk around black hole is proportional to the mass accretion rate, $\dot{M}$. This quantity can be assumed constant in the absence of outflows or convections. Integrating the mass conservation equation over polar angle, we find the net mass accretion rate, 
 \begin{equation}
 \dot{M}=-\int 2\pi r^2\sin\theta  \rho v_r  d\theta
 \end{equation} 
 Then we obtain,
 \begin{equation}
d\dot{M}/d\theta=-2\pi r^2 \rho v_r\sin\theta 
 \end{equation}
 Figure 8 displays the distribution of $d\dot{M}/d\theta$ along the $\theta-$direction. It's clear that $v_r$ is negative from the midplane and on the surface it becomes zero. In addition, $\rho$ is maximum in $\theta=90^\circ$, hence $d\dot{M}/d\theta|_{\theta=90}$ is positive and maximum. We can estimate the half of total mass accretion rate by calculating the area bounded between the graph of $d\dot{M}/d\theta$ and $\theta$ axis. As we see the larger poloidal magnetic field decreases the accretion rate, therefore we expect less luminosity from a flow in a stronger magnetic field. However, the asymmetric field makes accretion rate decrease in one side of the disk and increase in the other side of it. 
\begin{figure*}
\centering
\includegraphics[width=165mm]{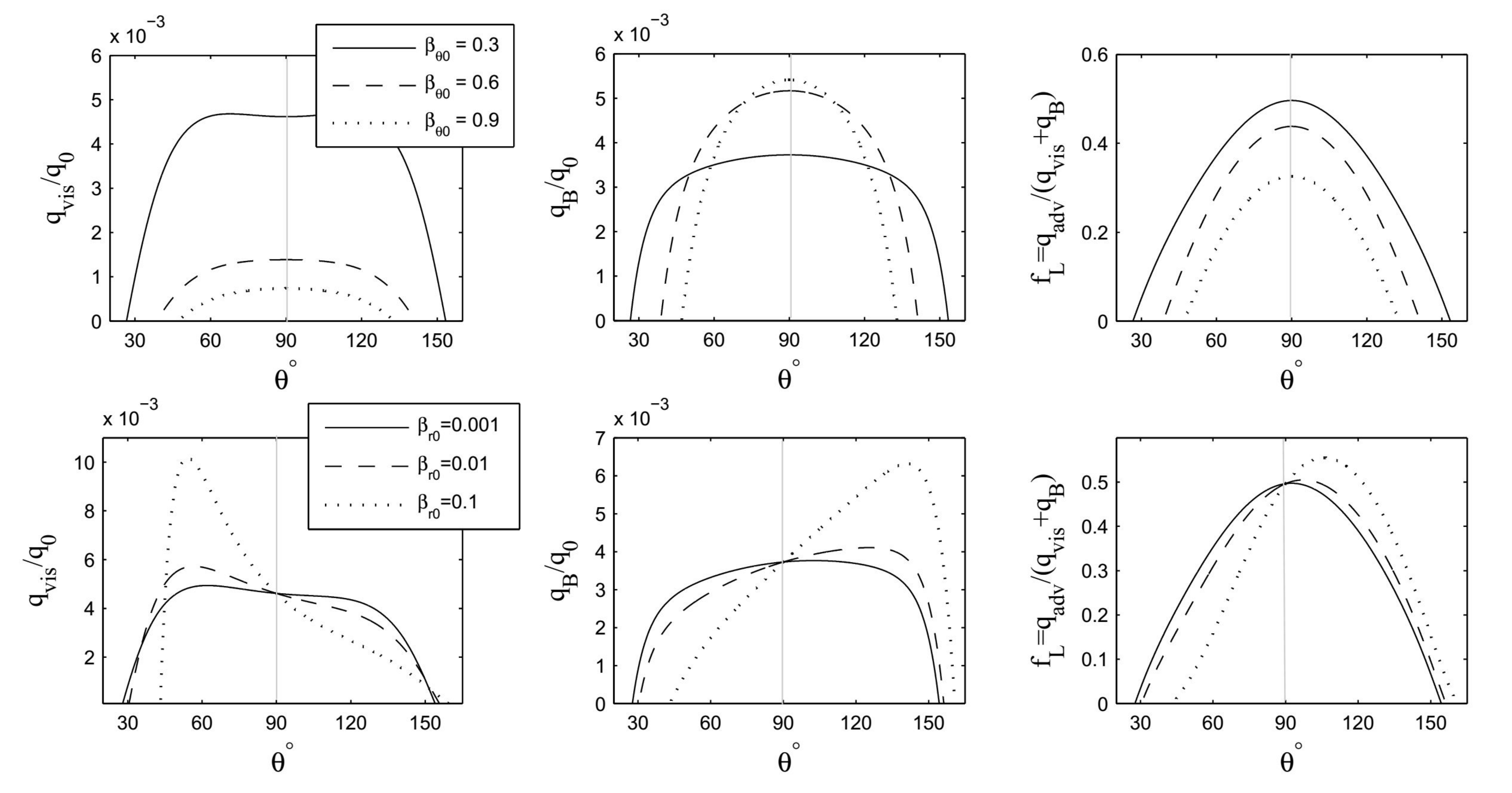}
 \caption{The latitudinal profile of the dissipated energies and local advection parameter. \textbf{Left side:} Dissipated energy  by viscosity  ($q_{vis}$) \textbf{Middle:}  Dissipated energy by magnetic resistivity, ($q_B$) per unit time and per unit volume of plasma, \textbf{Right side:} Local advection parameter $f_L$. Other parameters are  $\gamma=1.5$, $\alpha=0.1, K=1, Pr_m=1$, $\beta_{r0}=0$ in the case of symmetric magnetic field in the top panels and $\beta_{\theta0}=0.3$ for asymmetric fields in the bottom panels.  In this figure, we have scaled $q_{vis}$ and $q_B$ with a fiducial rate energy per unit volume which is defined by $q_0=p_K \Omega_K$.}
\end{figure*}

\subsection{Cooling and Heating Rates; Advection Parameter }
In hot accretion flows, we know turbulance viscosity is responsible to convert potential energy of gravity to thermal energy. Another source of heating in the magnetized flow is magnetic resistivity.  We can obtain both energies produced by viscosity and magnetic resistivity per unit time per unit volume as functions of polar angle, i.e. $q_{vis}(\theta)$ and $q_B(\theta)$. On the other hand, we know a certain fraction of total heating, i.e. $q_+=q_{vis}+q_B$, is transported to the disk's center, we show it as $q_{adv}=f_Lq_+$ and obviously it must vary in different parts of the disk with different polar angles. We can calculate the rate of heating per unit volume producing by 1) viscosity with this relation: $q_{vis}=\nu \rho r^2 (\partial\Omega/\partial r)^2$, and producing by 2) magnetic resistivity with $q_B= J^2/\sigma$ (notice $\sigma$ is the conductivity of plasma and it is inversely proportional to $\eta$, that is $\sigma=c^2/(4\pi\eta)$). Moreover, the rate of advected energy per unit volume can be determined by this formula: $q_{adv}=pv_r (\partial\ln p/\partial r-\gamma\partial \ln \rho/\partial r)/(\gamma-1)$. With self-similar assumption in radial direction, these quantities are simplified as
\begin{equation}
q_{vis}= \frac{9}{4}\frac{\alpha p v_\phi^2}{r v_K}
\end{equation} 
\begin{equation}
q_B= \frac{\eta}{4\pi r^2}\bigg(\frac{B_\theta}{4}+\frac{dB_r}{d\theta}\bigg)^2
\end{equation} 
\begin{equation}
q_{adv}= -\frac{5-3\gamma}{2(\gamma-1)}\frac{p v_r}{r}
\end{equation} 
The upper left side and middle panels of Fig.9 show similar behaviors of 2 ways of dissipation of energy inside the disk. Both viscous and magnetic energy dissipations are maximum at the equator. However, $q_{vis}$ is not noticable in a strong magnetic field especially at the equator. The second row of Fig.9  show the large value of $\beta_{r0}=0.1$ makes different aspect of $q_B$ and $q_{vis}$ in each side of the disk, so that one of them has a high peak near the surface and the other one is sharply decreasing there. For small values of $\beta_{r0}$, the heating rates are almost symmetric and show a mild slope in both sides. 
 
\begin{figure*}
\centering
\includegraphics[width=110mm]{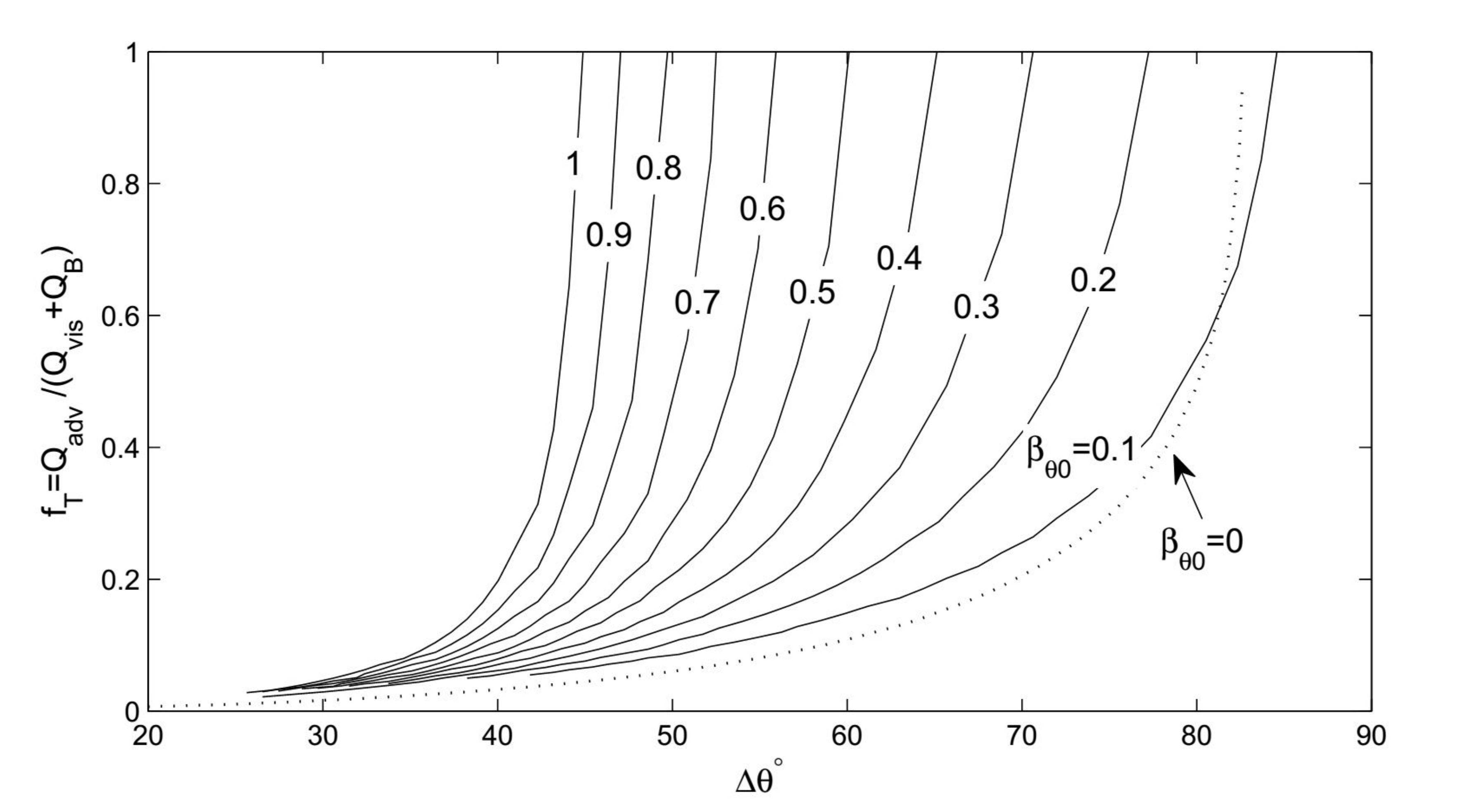}
\caption{Variation of the total advection parameter, $f_T$ with the disk's half-opening angle, $\Delta\theta$ corresponding to $\gamma=1.5$, $\alpha=0.1, Pr_m=1, \beta_{r0}=0$ and various value of $\beta_{\theta0}$.}
\end{figure*}

So far we have found density, pressure, velocity field and magnetic field components by solving mass, momentum conservation equations beside Faraday's induction equation. Here we determine two advection parameteres 1) local one, $f_L$, and 2) total one, $f_T$, which are specified by,
\begin{displaymath}
f_L=\frac{q_{adv}}{q_{vis}+q_B},
\end{displaymath}
\begin{displaymath}
f_T=\frac{Q_{adv}}{Q_{vis}+Q_B}=\frac{\int_{\theta_s}^{\pi-\theta_s} q_{adv}r\sin\theta d\theta}{\int_{\theta_s}^{\pi-\theta_s} q_{vis}r\sin\theta d\theta+\int_{\theta_s}^{\pi-\theta_s} q_Br\sin\theta d\theta}
\end{displaymath}
where $\theta_s$ is the surface angle and $q_i$ is local energy and $Q_i$ shows the summation of local energies in each vertical element of the flow. The local advection parameter (in the right side column of Fig.7) has maximum value at $\theta=90^\circ$ and becomes zero at the surface. In addition, less energy can be advected in each angle of the disk when the magnetic field is stronger. In the bottom right side of Fig.7, we see an asymmetric poloidal magnetic field doesn't change significantly the local advection parameter exception the fourth case with $\beta_{r0}=0.1$.

The total advection parameter for the symmetric disks is presented in figure 10. It shows the total advection parameter tends to unity in a thinner disk in a larger poloidal magnetic field.
 In other words, we can expect to see a fully advection dominated flow with slim appearance but if only the presence of strong large-scale field is plausible. Weak magnetic fields with $\beta_{\theta0}\leq 0$ which are common to use in simulations (and might be more real and stable) do not have appreciable effect on the thickness of ADAFs. 
Taking into account a purely toroidal magnetic field yields the same result on ADAFs (according to SAK14).

\section{Summary and Conclusion}
In this paper, we studied the vertical structure of axisymmetric viscous-resistive hot accretion flows with poloidal magnetic field. Following Gu et al. 2009, we neglected the meridional velocity ($v_\theta$) and assumed the dominant component of stress tensor is $T_{r\phi}$. 
Unlike Gu et al. 2009, we didn't need to employ the well-known polytropic relation of $p=K\rho^\gamma$ in the $\theta$-direction. Gu 2015 showed that optically thin accretion flows based on the polytropic assumption can not satisfy the thermal equilibrium, means that the advective cooling
is not able to balance the viscous heating. Moreover, Gu 2012 proved that polytropic
relation between the pressure and the density is not suitable to describe the vertical structure of radiation
pressure-supported disk. In our work, we found an extra equation from the adoption of $B_\phi=0$ in the induction equation and we could use it instead of the vertical polytropic equation.

Although the magnetic field generally depends on the dynamics of the flow, through the induction equation, we achieved an analytical relation for the magnetic flux. We divided the solutions into symmetric and asymmetric configurations. The self-similar approach led us a set of algebraic and differential equations with respect to polar angle. The proper boundary conditions were determined by the reflection symmetry assumption about the equator. We could find the amount of energy which advected towards the central compact object in each vertical element of the disk. As we expected the local advection parameter was maximum at the equator in the presence of the poloidal magnetic field with even symmetric configuration. The total advection parameter which is the same as one introduced by NY94 was found as a function of the half-thickness of the disk or opening angle $\theta_s$. Gu et al. 2009 had been pointed out that a fully advective regime leads an almost spherical accretion system, unlike them we have shown that for a strong magnetic field a fully advective disk would be expected for a slim shape. We clearly show that the stronger poloidal magnetic field helps the disk become advection dominance with smaller thickness. In other words, the symmetrical magnetic field whether poloidal (like this work) or toroidal (like SAK14) compress the disk. This effect is also seen in the different behaviors of gas pressure and magnetic pressure (the first one tends to expand the flow and the second one shrinks it). The asymmetric poloidal magnetic field was provided by taking into account a non zero radial component of the field at the equator. We saw with small ratio of $B_r/B_\theta$ at the equator, the disk roughly keeps its reflection symmetry about the mid-plane. Otherwise, remarkable differences happened in all quantities comparing one side with the other side of the disk.   

\section*{Acknowdlegment}
We are grateful to the anonymous referee for his/her thoughtful and constructive comments which helped us to improve this paper. MS also acknowledges Saramadan Federation for their kind supports to her projects.

\end{document}